\documentclass[preprintnumbers]{revtex4}
\usepackage{mathrsfs}
\usepackage{mathtools}
\usepackage{amstext}
\usepackage{hyperref}
\usepackage[none]{hyphenat}
\usepackage{color}

\usepackage{aas_macros}


\begin{document}

\title{Big-Bang Nucleosynthesis within\\ the Scale Invariant Vacuum Paradigm}
\author{V. G. Gueorguiev}
\affiliation{Ronin Institute, Montclair, NJ, USA}
\affiliation{Institute for Advanced Physical Studies, Sofia, Bulgaria}
\author{A. Maeder}
\affiliation{Geneva Observatory, University of Geneva, Switzerland}


\begin{abstract}
The Scale Invariant Vacuum (SIV) paradigm is applied to the Big-Bang Nucleosynthesis 
using the known analytic expressions for the expansion factor $a$ and
the plasma temperature $T$ as functions of the SIV time $\tau$ since the Big-Bang
when $a(\tau=0)=0$. The results are compared to the known standard
BBNS model as calculated with the PRIMAT code. Potential SIV-guided deviations from the 
local statistical equilibrium are explored. Overall, we find that smaller than usual baryon 
and non-zero dark matter content, by a factor of three to five times reduction, 
result in compatible to the standard reproduction of the light elements abundances.

\bigskip{}
\end{abstract}
\maketitle

\section{Introduction}\label{sec:intro}


It is often ignored that  when the cosmological constant $\Lambda$ is assumed to be equal to zero, 
the equations of General Relativity are scale invariant, 
which is also a property present in Maxwell equations of electrodynamics. 
However, for a non-zero $\Lambda$, the field equations of the gravitation no longer show the property of scale invariance.
A fact, which as discussed by \citet{Bondi90},  was one of the reasons of Einstein's disenchantment with the cosmological constant.
It is thus of interest to examine at what conditions the scale invariant properties of General Relativity may be restored,
since current cosmological observations support a  positive cosmological constant.
 
A theoretical framework has been developed by \citet{Dirac73} and \citet{Canu77}, 
in the so-called co-tensorial calculus based on the Weyl's Integrable Geometry (WIG) based on the original idea by \citet{Weyl23}. 
It offers a consistent basis to account for the properties of scale invariance of gravitation via a scale factor $\lambda$, 
as also illustrated by several properties studied by \citet{BouvierM78}.
Scale invariant derivatives, modified affine connexions, modified Ricci tensor and curvatures can be obtained  leading to a general scale
invariant field equation. Dirac and  Canuto et al. have  expressed an action principle in the scale invariant framework, 
with  a matter Lagrangian, as a coscalar  of power $n=-4$ (varying like $\lambda^{-4}$).
By considering the variations of this  action, they also obtain the generalization of the Einstein field equation. 
This is  Equation (7) in \citet{Maeder17a}  from  which the scale invariant cosmological equations  are derived. 
 
 In the Weyl’s Integrable Geometry (WIG),  the scale factor is undetermined without any other constraints
 as shown by Dirac and Canuto et al. Thus, these authors were fixing the scale factor by some external 
 considerations based on the so-called  Large Number Hypothesis; this hypothesis, however, is often disputed \citep{Carter79}. 
 Thus, it seems appropriate  to explore other conditions for setting the gauge.
 The proposition was made to fix the gauge factor by simply assuming  the  scale invariance of the empty space \citep{Maeder17a}. 
 This means that the properties of the empty space should not change under a contraction or dilatation of the space-time.
 Indeed, we note, as shown by  \citet{Carr92}, that the current equation of the vacuum 
 $p_{\mathrm{vac}}= - \varrho_{\mathrm{vac}} \, c^2$  already implies that    
 $\varrho_{\mathrm{vac}}$ should remain constant 
 ``if a volume of vacuum is adiabatically compressed or expanded''. 
 On this basis, the cosmological equations derived by \citet{Canu77} were simplified \citep{Maeder17a}.
 A number of cosmological tests were performed, with positive results. These equations were
 then shown to have rather simple analytical solutions  \citep{Jesus18} for models of a matter dominated Universe with a zero curvature.
 
In order to express the  motions of free particles, a geodesic equation was obtained  \citep{BouvierM78}
from a minimum action in the Weyl's integrable geometry (WIG). In the weak field approximation,
the geodesic equation leads to a  modification of the Newton equation \citep{MBouvier79} where it contains
a (currently very small) additional acceleration term proportional to the velocity of the particles. 
This equation was applied to study the internal motions of clusters of galaxies, 
the flat rotation curves of spiral galaxies and 
the age increase of the ``vertical'' velocity dispersion of stars in galaxies \citep{Maeder17c}.

The interesting result was that the  observational properties of these various systems
could be accounted without requiring to  the current hypothesis of dark matter, 
and the same for the radial acceleration relation (RAR) of galaxies \citep{Maeder18}. 
The growth of the density fluctuations in the early Universe was also studied by \citet{MaedGueor19}
who showed that dark matter is not needed, within the Scale-Invariant Vacuum (SIV) paradigm,
to achieve the growth of the density fluctuations to the currently observed inhomogeneities \citep{Coles02}.
Such studies suggested a connection between the Scale-Invariant Vacuum (SIV) theory and 
the origin of Dark Matter and Dark Energy \citep{MaedGueor20a}.
This was further reenforced by the study of scale-invariant dynamics of galaxies, 
MOND, dark matter, and the dwarf spheroidals \citet{MaedGueor20b}.
Furthermore, it was shown that the SIV framework naturally relates the scale invariance, horizons, and inflation,
while providing also a graceful exit from inflation \citet{SIV-Inflation'21}.
Summary of the main results was compiled and presented at the conference on
Alternative Gravities and Fundamental Cosmology (AlteCosmoFun'21) and published
in the journal Universe \citep{univ8040213}.

The above successes naturally inquire further studies as to the applicability of the SIV paradigm to other well-known phenomenon.
A study of the Cosmic Microwave Background \citep{Durrer08, Maeder17b} is one such phenomenon along with 
the Big-Bang nucleosynthesis (BBNS), which could be understood without a complicated numerical simulations 
\citep{Mukhanov04,Steigman07,Weinberg08}. The BBNS phenomenon is very relevant for us  
since the SIV possesses analytic expressions suitable for first exploration of such a problem \citep{Maeder18}.
This could be very useful as an approach to study BBNS within the SIV paradigm given 
by the recent Mathematica code PRIMAT \citep{PRIMAT}.
  
The objective of the present work is to apply the analytic expressions derived within the SIV paradigm \citep{Maeder18}
to the BBNS via the use of the PRIMAT code and to see how well the SIV will perform compared to the standard BBNS model. 
For this purpose, in Section \ref{sec:Background}, we provide a summary of the background needed by the reader to understand
the framework to be utilized. In Section \ref{sec:Method} are discussed the main methods, similarities and difference of various relevant functions, 
and the equations that need to be employed within the computational process. In Section \ref{sec:Results} we present our main results and 
explain the various model choices described in the tables shown. 
Finally, summary and conclusions are presented in the Section \ref{sec:Summary}.

\section{Background Framework}\label{sec:Background}

We start this section with a summary of the commonly used fundamental
physical constants and expressions relevant for the description of
the early Universe and their relation to the observations during the
current epoch:

\begin{eqnarray*}
H_{0} & = & h\,H_{100}\,,\;h=70/100\,,\;H_{100}=100\;\textrm{km/s/Mpc}=3.2408\times10^{-18}\,{\rm s^{-1}}\,,\\
\rho_{c0} & = & 3H_{0}^{2}/(8\pi\,G)\,,\;G=6.6743\times10^{-11}\,{\rm m^{3}/kg/s^{2}}\,,\;\tau_{0}=4.355\times10^{17}s\,,\;\\
T_{0} & = & 2.7255\,K\,,\;a_{BB}=\pi^{2}/(15\,\hbar^{3}\,c^{5})=2.31674\times10^{59}\,{s^{2}/m^{5}/J^{3}}\,,\;\\
k_{B} & = & 1.3806\times10^{-23}J/K\,,\;\rho_{\gamma0}=a_{BB}\left(k_{B}\,T_{0}\right)^{4}/c^{2}=4.6485\times10^{-34}\,{\rm g/cm^{3}}\,,\;\\
N_{\nu} & = & 3\,,\;K_{0}=1+\frac{7}{8}\left(\frac{4}{11}\right)^{4/3}N_{\nu}=1.6813\,,\;\rho_{\gamma0}\,h^{2}/\rho_{c0}=2.47476\times10^{-5}.
\end{eqnarray*}
Here, $H_{0}$ is the Hubble constant expressed via the reduced dimensionless
Hubble parameter $h$ and the scale fixing formal constant $H_{100}.$
The usual current critical density based on $H_{0}$ is $\rho_{c0}$, $G$ is Newton's gravitational constant, 
and $\tau_{0}$ is the current age of the Universe 
(13.8\,{\rm Gyr} with 365.25 days in a year as in \cite{Maeder18}).
Some minor differences from ref. \cite{Maeder18} are to be noted here:
the choice $h=0.7$ is used in ref. \cite{Maeder18} while PRIMAT uses Planck's
CMB value of $h=0.677$ \cite{PRIMAT} and
the pre-factor defining $v_{eq}$, in Eq. (27) of  \cite{Maeder18}, is $2.4741\times10^{-5}$
rather than the above value for $\rho_{\gamma0}\,h^{2}/\rho_{c0}$,
furthermore, the current value of the CMB temperature is $T_{0}=2.726\,K$ in  \cite{Maeder18},
and $a=a_{BB}k_{B}^{4}$ in Eq. (27) of ref. \cite{Maeder18}. 
PRIMAT \cite{PRIMAT} uses units such that $c=1$, $k_{B}=1$, and $\hbar=1$ with Planck's
CMB value of $h=67.66/100$ along with the number of effective neutrino flavors as $N_{\nu}^{eff}=3.01$,  
while $\Omega_{m}=0.31$ and $\Omega_{b}=0.05$ correspondingly.

The relevant SIV analytic expressions based on \cite{Maeder18}  are summarized
in the next set of formulas where the prefix ``A'' is used to indicate
that the subsequent equation number refers to the corresponding original equation
in ref. \cite{Maeder18}: 
\begin{eqnarray*}
(A27) & \; & v_{eq}=K_{0}\,\rho_{\gamma0}/(\Omega_{m}\rho_{c0})\,,\quad\; c_{2}=(v_{eq}^{2}+\sqrt{v_{eq}^{4}+C_{rel}})/t_{eq}^{2}\,,\;(A21)\\
(A20) & \; & C_{m}=4\Omega_{m}/(1-\Omega_{m})^{2},\,\;\quad C_{rel}=C_{m}\,v_{eq}\,,\;\\
(A25) & \; & t_{eq}=2^{-2/3}\left(v_{eq}^{3/2}(1-\Omega_{m})+\sqrt{v_{eq}^{3}(1-\Omega_{m})^{2}+4\Omega_{m}}\right)^{2/3},\\
(A29) & \; & t_{{\rm in}}=C_{rel}^{1/4}/c_{2}^{1/2}\,,\;\qquad\qquad\Delta{t}=(t_{0}-t_{{\rm in}})\,\tau/\tau_{0}\,,\;\qquad\qquad(A30)\\
(A33) & \; & a(\Delta{t})=\sqrt{2c_{2}t_{{\rm in}}^{3}\,\Delta{t}}\,,\;\qquad\tau(T)=\frac{T_{0}^{2}\tau_{0}}{2(t_{0}-t_{{\rm in}})\sqrt{C_{rel}}}\frac{1}{T^{2}},\quad(A39)\\
(A37) & \; & \rho_{r}(\Delta{t})=\rho_{\gamma0}\frac{K_{0}}{4C_{rel}\Delta{t}^{2}}\,,\;\quad\rho_{m}(\Delta{t})=\rho_{m0}\frac{c_{2}^{1/4}}{C_{rel}^{7/8}(2\Delta{t})^{3/2}}.
\end{eqnarray*}
where in (39) of \cite{Maeder18} one has $1.272\times10^{9}$ instead of $T_{0}\sqrt{\tau_{0}/2}=1.271\times10^{9}$ here.
The quantities $v_{eq}$ and $c_{2}$ are integration constants for the SIV modified FLRW equation and are related to the 
matter energy content $C_m$ and the radiation energy content $C_{rel}$, while  $t_{eq}$ is the moment of matter--radiation 
equality given in the SIV $t$--time such that the current time satisfies $t_0=1$. The moment of the Big-Bang (BB),
when $a=0$, is denoted by $t_{\rm in}\in \left[0,1\right)\,$, while $\Delta{t}$ is the time since the BB.  
$\Delta{t}$ is related via (A30) to the conventional time $\tau=\Delta\tau$ since the BB in seconds 
when $\tau_0$ is the current age of the Universe in seconds and the BB is at $\tau=0$. 
The expansion factor $a(\tau)$, also known as RW spatial scale factor, 
is given by substituting (A30) in (A33),
while (A39) gives the relationship between age $\tau$ and temperature $T$ of the radiation.
The last two expressions are the energy-densities for radiation and matter  within the SIV.

In PRIMAT, the thermonuclear reaction equations that describe the rate of change 
of the abundances of the various nuclear species are defined via in-scalar variables 
$Y_i=n_i/n_b$ based on the number density of the nucleus $i$ relative to  
the total baryon  number density $n_b$. 
For PRIMAT based expressions, we will use prefix P followed by the corresponding equation number in ref. \cite{PRIMAT}.
These equation numbers may differ by $\pm1$ between the arXiv version and the published version of the paper.
The usual reaction rates for the production and reduction of a specific nucleus
are re-expressed from the traditional form, i.e. a two body reaction $i+j\leftrightarrow k+l$ (P131), 
into the new form (P136), and also into a more general case of more bodies (P138), 
but the overall co-tensor structure stays the same since all $\Gamma$-s are now only in units of inverse seconds:
\begin{eqnarray*}
&(P131)&\dot{n}_i\supset n_kn_l\gamma_{kl\rightarrow{ij}}-n_in_j\gamma_{ij\rightarrow{kl}}\,,\;\qquad
\gamma_{ij\rightarrow{kl}}=\langle\sigma\,v\rangle_{ij\rightarrow{kl}}\,,\;(P132)\\
&(P136)&\dot{Y}_i\supset Y_kY_l\Gamma_{kl\rightarrow{ij}}-Y_iY_j\Gamma_{ij\rightarrow{kl}} \,,\;\qquad
\Gamma_{ij\rightarrow{kl}}=n_b\gamma_{ij\rightarrow{kl}}\,.\;\;\,(P137)
\end{eqnarray*}
Here, the reaction rate $\gamma_{j\ldots\rightarrow{i\ldots}}$ is in units  cm$^3$/s but when multiplied by 
the appropriate $n_b$ factor it results in $\Gamma_{j\ldots\rightarrow{i\ldots}}$ being in inverse seconds.

The forward $\gamma_{j\ldots\rightarrow{i\ldots}}$ and the reverse reaction rates 
$\bar{\gamma}_{j\ldots\rightarrow{i\ldots}}=\gamma_{i\ldots\rightarrow{j\ldots}}$
are related due to the assumption of a local thermodynamic equilibrium; 
thus, there is a simple three-parameter factor
containing the reaction constants $\alpha,\beta, \gamma$ and
expressed using temperature $T_9$ in GK units,
$\bar{\gamma}_{j\ldots\rightarrow{i\ldots}}=\gamma_{i\ldots\rightarrow{j\ldots}}
={\gamma}_{j\ldots\rightarrow{i\ldots}}\times\alpha\,T_9^\beta\,\exp{\left(\gamma/T_9\right)}$
(see (P141) and (P142) for details). 
The constant  $\alpha$ is an overall reaction constant related to the 
stoichiometric coefficients of the reaction species and their spins, 
while $\gamma$ is a factor in a Boltzmann like exponent and 
depends on the reaction Q-factor over a temperature constant of 1GK; 
as such $\sim\,Q/T$ is an in-scalar quantity if mass and thermal energy 
have the same $\lambda$-scaling.
Thus, the only co-scalar of non-zero power is related to the constant $\beta$
since it is coming from a factor of the type $m\times\,T$ (see (P141)). 
This means that if energy is scaling as $m\rightarrow\,m'=m\lambda^{n_m}$
and $k_BT\rightarrow\,k_BT'=k_BT\lambda^{n_T}$, 
then the effective $T_9$ re-scaling in the reverse reaction factor $T_9^\beta$
should be scaling as $\lambda^{(n_m+n_T)}$.
For most of our study we would assume that the scaling power of the rest-mass energy
and thermal energy are the same\footnote{
It is possible to argue for different scaling powers of the
radiation and rest-mass energies based on the different conservation laws
for matter ($w=0$) and radiation ($w=1/3$) based on the SIV conserved quantity
$\rho_w\,a^{3(1+w)}\lambda^{1+3w}=\rho_{0}$ within SIV. 
In doing so, one may induce a deviation from the usual energy conservation.
To avoid such deviation one will have to use the
appropriate equation of state  $w=p/\rho$ to determine the unique $\lambda$-scaling
for the energy.}, that is, $n_m=n_T$, otherwise it may result in apparent deviations 
from the law of energy conservation. Furthermore, we also adopt the
PRIMAT view that one can choose the system of units so that $k_B=1$
and therefore temperature is directly measuring thermal energy\footnote{
One can consider scaling for $k_B$, that is,  
$k_B\rightarrow\,k'_B=k_B\lambda^{n_{k_B}}$. 
However, since $k_B$ is a conversion constant from temperature to energy (erg/K),
as such it is related to the choice of units that once made should not be subject to change.
Thus, choosing $k_B=1$ fixes/eliminates this $\lambda$-scaling just like the the choice $c=1$
fixes the time and space units. 
However, one has to keep in mind whether the energy is related to thermal energy or rest-mass energy.},
thus there is no question of how the constant  $k_B$ scales with 
$\lambda$ since it is just a conversion constant that an observer 
can choose to be 1.

\section{Method}\label{sec:Method}

In PRIMAT one finds first the expansion factor $a(T)$, then
builds the energy density and its corrections as function of $a$
and/or $T$, finally the time variable $\tau$ is obtained 
from the FLRW equation\footnote{
This is a first order ordinary differential equation that needs
proper initial conditions. In this case it is set to be 
$\tau(a(T_i))=1/(2H(a(T_i))$ when integrating 
$d\tau/da=1/(a H(a))$ for $\tau(a)$, where $T_i\approx10^{12}$K is the 
initial temperature to be considered for the BBNS processes.}
$\dot{a}/a=H=\sqrt{\frac{8}{3}\pi\,G\rho}$. In
our approach to study BBNS within SIV, we bypass the numerical solution
of the FLRW equation in favor of using the analytic SIV
functions above. In particular, the functions used in the SIV-PRIMAT
are: $a(T)=a(\tau(T))$ based on $\tau(T)$ and $a(\tau)$ above,
while the inverse function $T(a)$ is computed within PRIMAT.
\footnote{
The PRIMAT (numerical) inverse function process is validated by 
comparing the numerically inverted function to the known SIV analytic function $T(a)$.
We use the same validation approach as done by the original PRIMAT code for other similar cases.}
This way densities are not needed for FLRW integration to obtain 
$\tau(a)$ and $a(\tau)$.

\begin{figure}[th]
\includegraphics[width=0.9\textwidth]{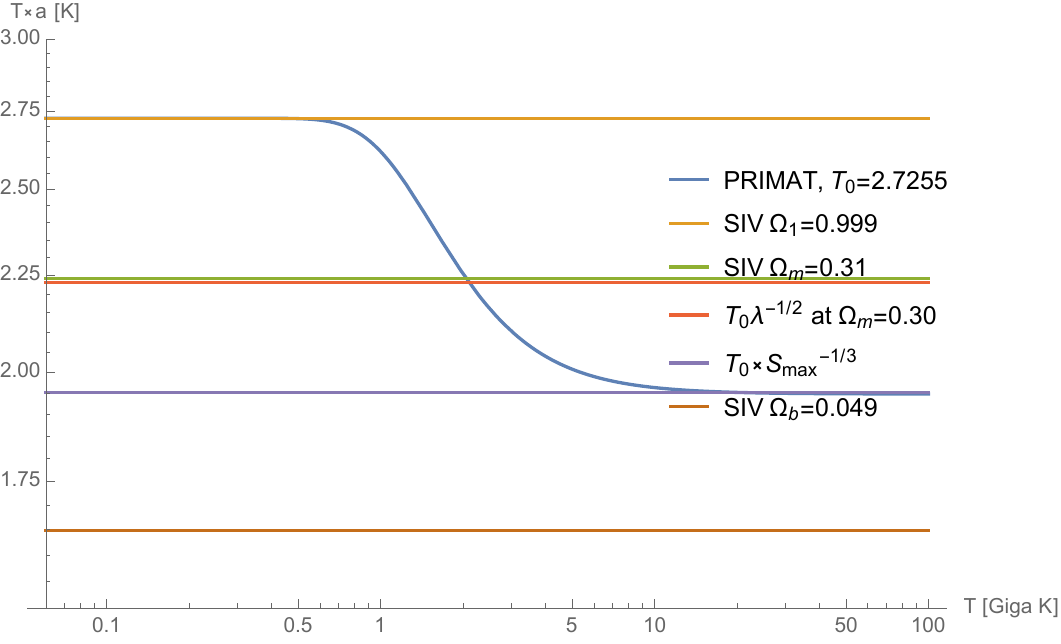}
\caption{\small{$T\times a(T)\rightarrow T_{0}$ for $\Omega_{m}=0.999$ ($\lambda=1$)
coincides with PRIMAT's function at low temperatures ($T<1GK$). This
is in agreement with $\lambda=1/t_{{\rm in}}\rightarrow1$ as $t_{{\rm in}}\rightarrow1$
while $\Omega_{m}\rightarrow1$. At high-temperatures ($T>1GK$) PRIMAT
$a(T)$ is smaller than the SIV for $\Omega_{m}=0.31$, but bigger
than the SIV case of $\Omega_{m}=\Omega_{b}=0.05$. 
PRIMAT is running with $\Omega_{m}=0.31$
and $\Omega_{b}=0.05$; that is, $\Omega_{CDM}=0.26$, but this value
or any other nearby value do not have an impact on the BBNS within
PRIMAT! The high-temperature ($T>1GK$) regime in PRIMAT is due to
the neutrino physics and $e^+e^-$ annihilation.
The high-temperature limit of $T\times a(T)$ for PRIMAT is based on the 
numerical value of  the distortion factor 
$\mathscr{\mathcal{S}}(T)\rightarrow\mathscr{\mathcal{S}}_{max}=11/4\approx\,T_0$ at $T\gg1GK$.
In order to distinguish similar color curves notice that the top-to-bottom labeling in the legend 
corresponds to the curves ordering at low temperatures.}}
\label{T*a(T)} 
\end{figure}

As usual, the functional form of the expansion factor $a$ is
inversely proportional to the temperature $T$. As can be seen from (A39),
(A30), and (A33) $a(T)=a(\tau(T))=\,const/T$ where the constant is
$T_{0}\,C_{{\rm rel}}^{-1/4}\sqrt{c_{2}t_{{\rm in}}^{3}}$ which by
(A29) becomes $T_{0}\sqrt{t_{{\rm in}}}=T_{0}\lambda^{-1/2}$. This
constant depends only on $\Omega_{m}$ and the CMB temperature $T_{0}$.
During the BBNS $\lambda$ is practically constant since it is
very close to $1/t_{\rm in}$. However,  $\lambda$ is generally
evolving during the evolution of the Universe towards the
value $\lambda_{0}=1$ at the current epoch. In the case of $\Omega_{m}\rightarrow1$
one also has $\lambda\rightarrow1.$ Either way, one obtains $a\rightarrow T_{0}/T\rightarrow1$
towards the current epoch. In PRIMAT for neutrino decoupled scenario
the expansion factor $a(T)$ is also of the form $T_{0}/T$ but has a
distortion factor of $\mathscr{\mathcal{S}}(T)^{-1/3}$ due to 
neutrino physics and entropy injection by the electron-positron annihilation
process around $T\sim1GK$. The distortion factor $\mathscr{\mathcal{S}}(T)$
becomes 1 for $T\ll1GK$ and therefore recovers the usual $a(T)=T_{0}/T$
behavior in the low-temperature regime (see Fig. \ref{T*a(T)} for details).
Note that if one interprets $\Omega_{m}$ and $\Omega_{b}$ as the current epoch (now)
values, then $\Omega_{m}=1$ is not realistic limit here given the known current values; 
however, since the BBNS is in the radiation epoch far from the matter-radiation equality
when radiation dominates then $\Omega_{m}=1$ for the total matter and radiation seems reasonable;
thus, we consider $\Omega_{m}=1$ along with $\Omega_{m}=0.3$ and $\Omega_{m}=\Omega_{b}=0.05$
to illustrate the trend due to $\Omega_{m}$ in the graphs.

To define the time variable, PRIMAT solves the FLRW equation
using the relevant energy density. Within the SIV we have an analytic
expressions for the SIV time. In the standard units 
(age of the Universe is $13.8\,{\rm Gyr}$), the SIV analytic form is
$\tau(a)=const\times\,a^{2}$ and from (A30) and (A33) the constant is:

\begin{eqnarray*}
\tau(a)/a^{2} & = & \tau_{0}/\left(2c_{2}\,t_{{\rm in}}^{3}(t_{0}-t_{{\rm in}})\right)
=\tau_{0}/\left(2C_{rel}^{1/2}\,t_{{\rm in}}(t_{0}-t_{{\rm in}})\right)=\\
& = &\tau_{0}(1-\Omega_{m})/\left(4\sqrt{\Omega_{m}\,v_{eq}}\,t_{{\rm in}}(t_{0}-t_{{\rm in}})\right)
=\tau_{0} f(\Omega_{m}).
\end{eqnarray*}
This constant depends only on $\Omega_{m}$ and for
$\Omega_{m}\rightarrow1$ goes to $3\tau_{0}/(4\sqrt{\,v_{eq}(\Omega_{m}=1)})$.

\begin{figure}[th]
\includegraphics[width=0.9\textwidth]{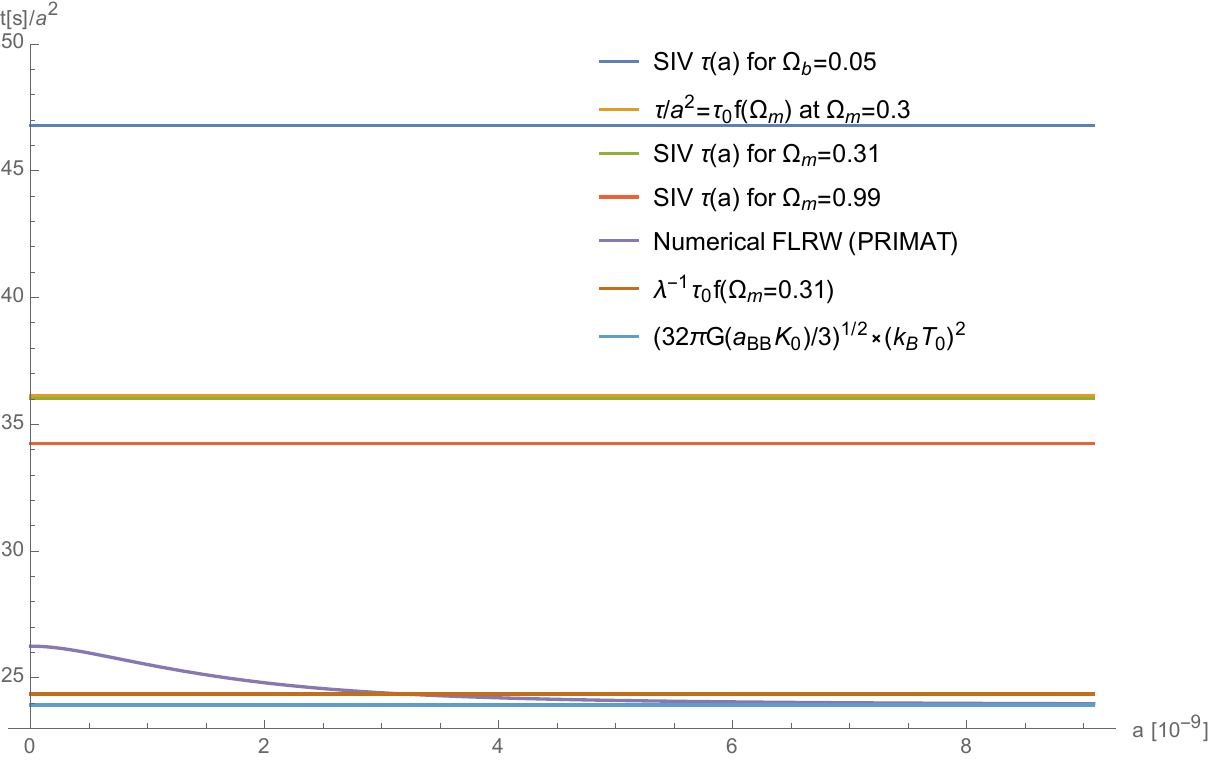}
\caption{\small{
The SIV constant $\tau(a)/a^{2}$ 
decreases with $\Omega_{m}$ but aways stays above the PRIMAT value.
The gap between $\Omega_{m}=1$ ($\lambda=1$) and the PRIMAT $\tau(a)$ can be resolved
by the $\lambda$--scaling of $8\pi\,G\rho$ as seen in the bottom curves. 
The bottom line is the asymptotic limit of $\tau(a)/a^{2}$ for PRIMAT
based on the integration of $\sqrt{8\pi/3\,G\rho_{\gamma0}}$ using
the low-temperature radiation density limit for PRIMAT density $\rho$.
The label on the vertical axes is set to be $t[s]/a^2$ in order to remind us that
within PRIMAT time $t$ is in seconds; thus, this $t[s]$ is not the SIV dimensionless time $t\in[0,1]$,
but actually it is the $\tau[s]$ within the SIV since it is in seconds since the BB.
Top-to-bottom labeling in the legend 
corresponds to the curves order at low temperatures.}}
\label{tau(a)} 
\end{figure}

As it is seen in Fig.~\ref{tau(a)}, during the relevant interval of time, 
there is a clear quadratic relation of $\tau(a)\,\sim\,a^{2}$
that becomes obvious on the $\tau(a)/a^{2}$ plot. In the limit
$\Omega_{m}\rightarrow1$, one has $C_{m}$, $C_{rel}$, and $c_{2}\rightarrow\infty$
in this respect $t_{eq}$ is sandwiched between $t_{{\rm in}}$ and
$t_{0}=1$ and based on (A25) $t_{eq}\rightarrow1$. Notice that the
PRIMAT $\tau(a)/a^{2}$ is larger in the initial stages of the BBNS
and then becomes smaller then the SIV $\tau(a)/a^{2}/\lambda$ but
it is about the same order of magnitude. 
The $\lambda$--scaling is due to the corresponding $\lambda^{2}$--scaling
of the $8\pi\,G\rho$ within the SIV and the fact that the PRIMAT
time is based on integrating the FLRW equation $\dot{a}/a=\sqrt{\frac{8}{3}\pi\,G\rho}$.
The time keeping, between the PRIMAT and SIV,  is non-uniform (as seen in Fig. \ref{tau(a)}); 
this has an impact on the overall time related observables, such as life-time of processes and particles, 
i.e. neutron life-time and nuclear reaction rates within the SIV framework.
The details of the PRIMAT $\tau(a)/a^{2}$ variations
could be understood to be due to the high-temperature behavior of the relativistic density 
containing correction terms $\delta\rho(T)$ and also via the dependence on $a(T)$,
which in PRIMAT has the distortion factor $\mathscr{\mathcal{S}}(T)$ 
that is missing in the SIV model. 
As it can bee seen $\tau(a)/a^{2}$  decreases with $\Omega_{m}$ 
but  always stays above the PRIMAT value.
The gap between $\Omega_{m}=1$ ($\lambda=1$) 
and the PRIMAT $\tau(a)$ can be resolved by the 
$\lambda$--scaling of $8\pi\,G\rho$ as seen in the bottom curves. 
This means that if one uses the SIV $a(T)$ instead of the default PRIMAT functions,
but utilizes the PRIMAT densities and usual FLRW to define $\tau(a)$
then one would obtain a curve similar to the one displayed but pushed 
by a factor $\lambda$  to the corresponding SIV $\tau/a^2$ 
line shown at  $\Omega_{m}=0.3$. The reason for this is the 
factor  $\lambda$ in the SIV $T\sim\lambda^{-1/2}$, which propagates 
through the density $\rho\sim\,T^4$ to $\lambda^{-2}$ factor, that 
becomes $\lambda^{-1}$ due to the square root in $\dot{a}/a=\sqrt{\frac{8}{3}\pi\,G\rho}$,
and ultimately resulting in $d\tau'=\lambda\,d\tau$, which is consistent 
with the SIV view about its effects on time and space intervals.

\begin{figure}[th]
\includegraphics[width=0.9\textwidth]{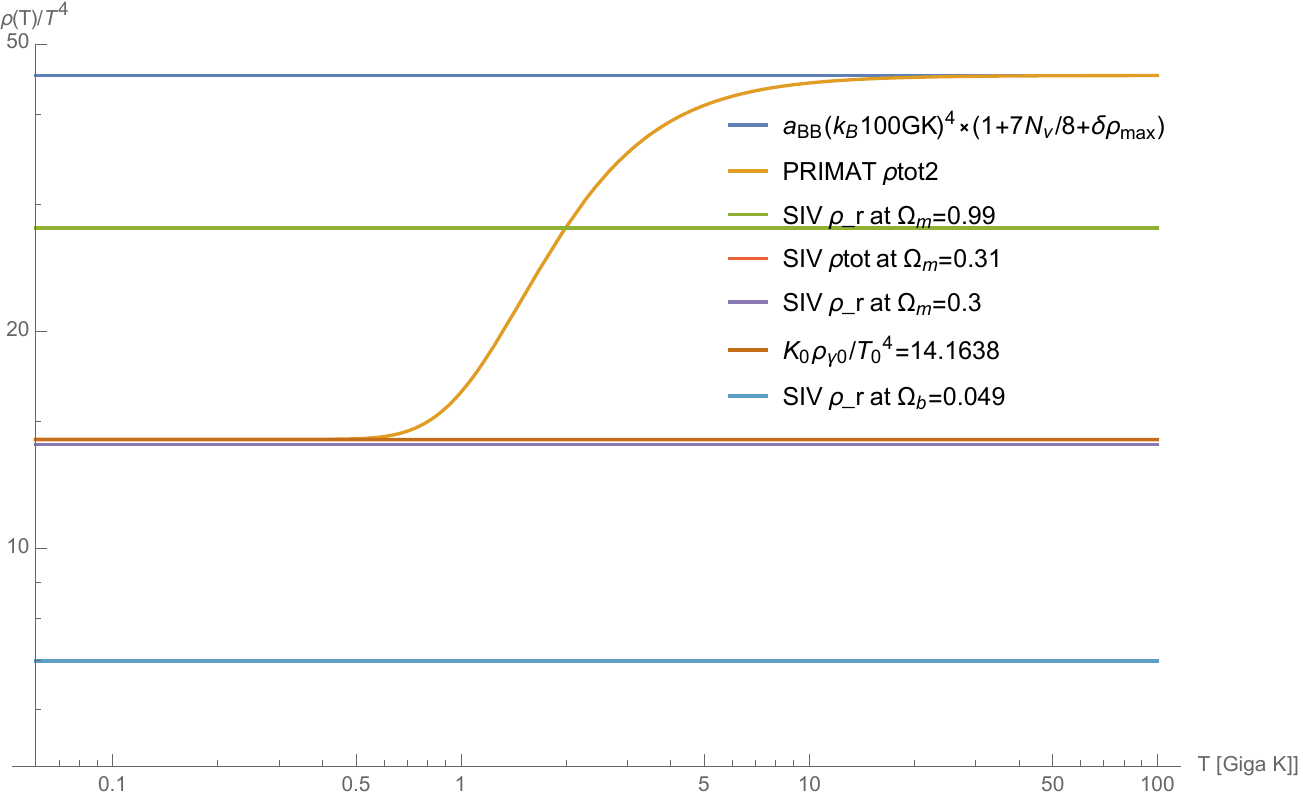}
\caption{\small{The ratio $\rho(T)/T^{4}$ for SIV ($\rho_{tot}$ and radiation
$\rho_{r}$) at $\Omega_{m}=0.999$ are below the PRIMAT $\rho(T)$
function at high temperatures ($T>1GK$) due to the plasma corrections
in this regime. The top line represents the high-temperature limit as
evaluated at $N_\nu=3$ and $\delta\rho_{max}\approx1.742$ at $T=100GK$.
At low-temperatures ($T<1GK$) PRIMAT $\rho(T)$ coincides
with the SIV $\rho_{tot}$ for $\Omega_{m}=0.31$. It matches the
value $K_{0}\rho_{\gamma0}/T_{0}^{4}$. Note that $\Omega_{m}=1$
is not realistic limit here since the BBNS is in the radiation epoch but can be viewed 
as another way to say $\lambda=1$. There are three lines near the ratio of 14 
and practically on top of each other. Top-to-bottom labeling in the legend 
corresponds to the curves order at high temperatures.}}
\label{rho(T)} 
\end{figure}
Below are the densities used in PRIMAT for the case of decoupled neutrinos: 
\begin{eqnarray*}
\rho_{\gamma} & = & a_{BB}\left(k_{B}T\right)^{4}/c^{2}
\left(1+\delta\rho(T)+\frac{7}{8}N_{\nu}\left(\frac{\left\langle T_{\nu}\right\rangle }{T}\right)^{4}\right)
=T^{4}\overline{\rho}_{\gamma},\\
\rho_{m} & = & \frac{n_{b0}m_{b0}}{c^{2}a^{3}}
\left(1+\Omega_{c0}/\Omega_{b0}+\frac{3}{2}k_{B}T/m_{b0}\right)
=\frac{a_{0}^{3}\rho_{m0}(T)}{a^{3}}.
\end{eqnarray*}
In the high-temperature regime $\rho_{tot}\sim\,a_{BB}\left(k_{B}T\right)^{4}/c^{2}$.
Where the proportionality constant is $\left(1+\delta\rho_{max}+7N_{\nu}/8\right)$
since at high-temperature the neutrinos are in thermal equilibrium with
the radiation; that is $\left\langle T_{\nu}\right\rangle =T$. In
the low-temperature regime this is $K_{0}$ since the plasma corrections
$\delta\rho$ are becoming negligible
and $\left\langle T_{\nu}\right\rangle /T\rightarrow(4/11)^{1/3}$
see PRIMAT (P32). This is demonstrated in Fig.~\ref{rho(T)}.

For the rate of change of the nuclear species, 
we have to consider that the right hand side of the equation (P136) is in SIV time $\tau$
while the original PRIMAT reaction rates $\Gamma'_{j\ldots\rightarrow{i\ldots}}$
are in the Einstein GR frame, 
we will use a prime to indicate that, but these rates need to be expressed in the SIV frame.
Based on (P136) we have for the transition from EGR to WIG (SIV) frame:
\[
\frac{dY_i}{d\tau'}=Y_kY_l\Gamma'_{kl\rightarrow{ij}}-Y_iY_j\Gamma'_{ij\rightarrow{kl}},\;\Rightarrow
\frac{1}{\lambda}\frac{dY_i}{d\tau}=Y_kY_l\Gamma'_{kl\rightarrow{ij}}-Y_iY_j\Gamma'_{ij\rightarrow{kl}}.\;
\]
Thus, $\Gamma_{kl\rightarrow{ij}}$ in the SIV frame is related to the measured EGR laboratory rates 
$\Gamma'_{kl\rightarrow{ij}}$ via a simple rescale factor $\lambda$; that is,
$\Gamma_{kl\rightarrow{ij}}=\lambda\Gamma'_{kl\rightarrow{ij}}$, which is
based on the relationship\footnote{Even though 
$d\tau'$ in EGR is in seconds and so is the WIG time interval $d\tau$, 
these two time units (seconds) are not necessarily the same. If they were the same, 
then the EGR and WIG frames would coincide since this would imply $\lambda=1$.
Furthermore, the relation $d\tau'=\lambda\,d\tau$
could be viewed as a consequence of the $\lambda$ scaling of $\rho$ and the definition of 
$\tau$ via the solution of the FLRW equation as it was discussed in connection to the $\rho/T^4$ relationship.}
$d\tau\rightarrow\,d\tau'=\lambda\,d\tau$; 
thus, the original PRIMAT rates need to be rescaled as
$\Gamma'_{j\ldots\rightarrow{i\ldots}}\rightarrow\,\Gamma_{j\ldots\rightarrow{i\ldots}}
=\lambda\times\Gamma'_{j\ldots\rightarrow{i\ldots}}$.
This is accomplished by using Forwards Rescale Factor (FRF) to all the reaction rates. 
That is, FRF=$\lambda$ for SIV guided studies\footnote{
One can argue that FRF should be 1 because the reaction 
cross-sections $\sigma$ should not be modified since the sizes of the nuclei are not governed by gravitation.
Such argument ignores the possibility that $\hbar$ may not be an in-scalar object.
Nevertheless, we can carry on and consider 
$\dot{n}_i\supset\,n_i\,n_j \langle\sigma\,v\rangle_{ij\rightarrow{kl}}$ based on (P131) and (P132).
Thus, because $v$ is inscalar, there is no change on the RHS of the equation. 
However, our argument was about the change of the time parametrization 
in the LHS that takes into account $d\tau\rightarrow\,d\tau'=\lambda\,d\tau$.
So, the $\lambda$ in the denominator of the LHS becomes a FRF scale-factor on the RHS
when one is using a different time parametrization in switching from EGR to WIG (SIV) formulation.}. 
Furthermore, the $T_9$ argument of the factor $T_9^\beta$ in the reverse reaction rates
may have to be rescaled as well with the appropriate $\lambda$-factor 
for mass and thermal energy scaling. 
That is, for SIV guided studies, 
$mT_9\rightarrow\,\text{mT}_9\times\lambda^{n_m+n_T}$. 
In our results section, we refer to the scale-factor $\lambda^{n_m+n_T}$
as \text{m\v{T}} scale-factor $\lambda^{2n}$ when $n_m=n_T=n$.
However, the two scale-factors powers $n_m$ and $n_T$ may not be the same,
but we have deferred the discussion on this topic to the Appendix \ref{sec:Appndx}.
 
In this paragraph we drop the primes since the discussion is about the standard BBNS.
The validity of  $Ta=T_0a_0$ is affected during the standard BBNS due to the $e^+e^-$ annihilations, 
as such it is related to the distortion factor $\mathscr{\mathcal{S}}(T)$ (see P31);
that is, $a_0T_0=aT\mathscr{\mathcal{S}}^{1/3}(T)$. Furthermore,
since $n_b\propto\,a^{-3}$ and $n_\gamma\propto\,T^3$,
this also affects the baryon to photon in-scalar ratio 
$\eta=n_b/n_\gamma=\eta_0\mathscr{\mathcal{S}}^{1/3}(T)$,
where $\eta_0=6.0913\times10^{-10}$ is the current ratio of  baryons to photons;
it is often written as  $\eta_{10}=\eta_0\times10^{10}$ which removes the factor  $10^{-10}$ in  $\eta_0$.

Finally, we would like to point out that if the SIV paradigm is valid, and since during the BBNS $\lambda$ 
is practically constant, then one could include the effect of the $e^+e^-$ annihilation via the distortion factor 
$\mathscr{\mathcal{S}}(T)$ as an equivalent effect within the SIV background.
To do so, we recognize that for EGR $\leftrightarrow$ SIV with $a'=a\lambda$ and $T'=T\lambda^{-1/2}$, 
one has $a_0T_0(\mathscr{\mathcal{S'}}(T'))^{-1/3}=T'a'=Ta\lambda^{1/2}$ and therefore the new
$\tilde{a}(T)=a_0T_0/(T\lambda^{1/2})/\mathscr{\mathcal{S}}^{1/3}(T)=a_{SIV}(T)/\mathscr{\mathcal{S}}^{1/3}(T)$,
where $\mathscr{\mathcal{S}}(T)$ is defined according to the discussion in the Appendix \ref{sec:Appndx}  
via the known EGR laboratory function $\mathscr{\mathcal{S}}'(T')$, that is, 
$\mathscr{\mathcal{S}}(T)=\mathscr{\mathcal{S}}'(T'(T))=\mathscr{\mathcal{S}}'(T\lambda^{-1/2})$.
Furthermore, the new $\tilde{a}(T)$ is also equivalent to $a'(T')/\lambda$,
as it should be based on $a'=a\lambda$.

As a measure of goodness of fit for the theory against the experimental data,
we use  $\sqrt{\chi_{\epsilon}^{2}}$, which is the 2-norm of
the deviation of theory (th) from observation (ob) with respect to
the experimental error $\epsilon$.
\[
\sqrt{\chi_{\epsilon}^{2}}=\sqrt{\frac{1}{N}\sum_{i}^{N}
\left(\frac{y_{i}^{(ob)}-y_{i}^{(th)}}{\epsilon_{i}}\right)^{2}}.
\]
A number less than one indicates that all the theory
values are within the observational uncertainties. 

\begin{table}[h]
\small
\begin{tabular}{||c||c|c||c|c||c|c||}
\hline \text{T [GK]} & \text{a(T)\;}[$10^{-9}$] & \text{SIV a(T)} & \text{\ensuremath{\tau} [s]} 
& \text{SIV \ensuremath{\tau}[s]} & $\rho_{\text{tot2}}$ & \text{SIV}$\rho_{\text{rad}}$\\
\hline 9 & 0.218434 & 0.249208 & 1.24909 & 2.23615 & 287845. & 92875.6\\
6 & 0.33151 & 0.373813 & 2.86926 & 5.03133 & 54801.9 & 18345.8\\
5 & 0.401382 & 0.448575 & 4.19779 & 7.24512 & 25683.6 & 8847.32\\
4 & 0.509668 & 0.560719 & 6.7449 & 11.3205 & 9997.75 & 3623.86\\
3 & 0.700877 & 0.747625 & 12.6701 & 20.1253 & 2856.65 & 1146.61\\
2 & 1.12735 & 1.12144 & 32.2898 & 45.282 & 445.583 & 226.491\\
1 & 2.61734 & 2.24288 & 168.002 & 181.128 & 16.4666 & 14.1557\\
0.9 & 2.95001 & 2.49208 & 212.491 & 223.615 & 10.2532 & 9.28756\\
0.8 & 3.35699 & 2.80359 & 274.024 & 283.013 & 6.1356 & 5.79818\\
0.7 & 3.86726 & 3.20411 & 362.272 & 369.649 & 3.49112 & 3.39879\\
0.5 & 5.44844 & 4.48575 & 714.557 & 724.512 & 0.887353 & 0.884732
\\\hline 
\end{tabular}
\caption{\small{Values of $a(T)$, $\tau(T)$, and $\rho(T)$ for PRIMAT and SIV 
using standard cosmological parameters: 
$\Omega_{CDM}=0.26$, $\Omega_{b}=0.05$, and $h=0.677$.
The PRIMAT $\rho_{\text{tot2}}$, corresponds to the densities discussed earlier,
uses an effective number of neutrino flavors $N_\nu=3.01$.
The densities $\rho$ are in g/cm$^3$.}}
\label{Table1} 
\end{table}

\section{Results}\label{sec:Results}
In Table \ref{Table1} we have shown the values of $a(T)$, $\tau(T)$, and
$\rho(T)$ for PRIMAT when using standard cosmological parameters
for $\Omega_{CDM}=0.26$ and $\Omega_{b}=0.05$ along with
the corresponding values for the relevant SIV functions for the same
cosmological parameters\footnote{
We have used temperature T as the 
control variable, which is customary; 
however, an in-scalar quantity will be more appropriate
since there could be a $\lambda$-scaling for T within SIV.}. 
From Table \ref{Table1} and from Fig.  \ref{tau(a)} we see that the two ``clocks''
are irregular in the few first moments after the Big-Bang 
with SIV time ticking about 1.008 times faster than the PRIMAT time;
however, at low temperatures they become synchronized
and shifted only by a few seconds.
This may be taken as a justification to use FRF=1 since 
the 1.008  is practically 1.

\begin{table}[h]
\small
\begin{center}
\begin{tabular}{|c|cc|cc|cc|}
\hline\hline 
Element & Obs. & PRMT & $a_{SIV}$ & fit & $\bar{a}/\lambda$ &  fit* \\
\hline 
\text{H} & 0.755 & 0.753 & 0.805 & 0.755 & 0.75 & 0.753 \\
$Y_P=4Y_{\text{He}}$ & 0.245 & 0.247 & 0.195 & 0.245 & 0.25 & 0.247  \\
$\text{D/H}\times10^5$ & 2.53 & 2.43 & 0.743 & 2.52  & 1.49 & 2.52 \\
\hline  
$^3\text{He/H}\times10^5$ & 1.1 & 1.04 & 0.745 & 1.05 & 0.884 & 1.05 \\
$^7\text{Li/H}\times10^{10}$ & 1.58 & 5.56 & 11.9 & 5.24 & 9.65 & 5.31 \\
\hline  
$N_{\text{eff}}$ & 3.01 & 3.01 & 3.01 & 3.01 & 3.01 & 3.01 \\
$\eta_{10}$ & 6.14 & 6.14 & 6.14 & 1.99 & 1.99 & 5.57  \\
$\Omega _b$\; [\%] & 4.9 & 4.9 & 4.9 & 1.6 & 1.6 & 4.4 \\
$\Omega _m$\;[\%]  & 31 & 31 & 31 & 5.9 & 5.9 & 86 \\
$\sqrt{\chi _{\epsilon}^2}$ & N/A & 6.8 & 35 & 6.1 & 22 & 6.2  \\
\hline\hline 
\end{tabular}
\end{center}
\caption{\small{
Abundances of the light elements within the standard (PRIMAT) and SIV BBNS. 
The observational uncertainties are  1.6\% for $Y_{P}$, 1.2\% for D/H, 18\% for He/H, and 19\% for Li/H.
The column denoted by fit contain the results for perfect fit 
on $\Omega _b$ and $\Omega _m$  to $^{4}$He and D/H,
while fit* is the best possible fit on $\Omega _b$ and $\Omega _m$ 
to the $^{4}$He and D/H observations 
for the model considered as indicated in the columns four and six.
The last four columns depict results produced by 
using the analytical functions within a ``naive'' SIV model without modification of any of the reaction rates.
The last two columns are usual PRIMAT runs with modified $a(T)$ 
such that $\bar{a}/\lambda=a_{SIV}/\mathcal{S}^{1/3}$,
where $\bar{a}$ is the PRIMAT's $a(T)$ for the decoupled neutrinos case.
Column six is actually $a_{SIV}/\mathcal{S}^{1/3}$,
but it is denoted by $\bar{a}/\lambda$ to remind us 
about the relationship $a'=a\lambda$;
the run is based on $\Omega _b$ and $\Omega _m$ from column five.
The different values of $\eta_{10}$ are due to different values of
$h^2\Omega_b$, which can be seen by noticing that 
$\eta_{10}/\Omega_b$ is always $\approx 1.25$.}}
\label{Table2} 
\end{table}

The relevant element abundances are given in Table~\ref{Table2}. 
In the second column are shown the observational values,
while in the third column are the results of the PRIMAT code
when run with small reaction network and decoupled neutrinos,
without QED dipole corrections\footnote{When the partially decoupled neutrinos 
and QED dipole corrections are turned on there are minor insignificant changes
to the results that are not relevant to the current discussion.}, 
with standard $\Omega _b$ and $\Omega _m$ values. 
The forth column shows the SIV results for the same values of 
$\Omega _b$ and $\Omega _m$ using the 
analytic SIV functions $a_{SIV}(T)$ and $\tau_{SIV}(T)$.
The results reveal under-production of 
$^4$He, deuterium, and $^3$He with significant 
over-production of $^7$Li. Abundances improve if we fit the
$^4$He and D/H by changing the values of
$\Omega _b$ and $\Omega _m$ as seen in the fifth column.
Now $^3$He and $^7$Li are compatible with the PRIMAT results,
but at much smaller values of the baryon and total matter. 
In this case the dark matter (DM) is less than $3\times$ of the baryon matter (BM),
unlike the usual case where the DM is more than $5\times$ of the BM. 

In order to study the contribution of $\mathscr{\mathcal{S}}(T)$  within SIV
we consider the last two runs (columns six and seven in Table \ref{Table2}).
The sixth column is based on the parameters in the fifth column.
Not shown in the Table, but if we use the PRIMAT values for $\Omega _b$ and $\Omega _m$  (based on column three) 
instead of column five, then we have over-production of $^4$He and  significant under-production of deuterium and 
very high production of $^7$Li - all in the further enhanced directions of the results in column four.
So, the last two columns are for $a_{SIV}(T)$ distorted by $\mathscr{\mathcal{S}}(T)$ 
which is equivalent to PRIMAT $\bar{a}(T)$ distorted by $\lambda$.
Such choice of modification is relevant since 
the SIV runs (column four and five) don't include the 
electron--positron annihilation and neutrino effects
encoded in the function $\mathscr{\mathcal{S}}(T)$.
Adding the distortion function $\mathscr{\mathcal{S}}(T)$
to the $a_{SIV}(T)$ or equivalently modifying PRIMAT $\bar{a}(T)$  by $\lambda$, 
results in light increase of $^4$He and under production of deuterium and tritium
with an over production of $^7$Li (comparing column six to five). 
The next, seventh column, fit* is the best possible, but not perfect, 
fit for $^4$He and D/H and seems to require a significant mass content.
This is a simple (``naive'') SIV model without the utilization of the $\lambda$-modifications of the 
various reaction parameters (FRF,  \text{m\v{T}},  \text{Q/\v{T}}) 
that have been discussed in the Appendix \ref{sec:Appndx}. 
Thus, this fit* failure to achieve a perfect 2D fit (on $^4$He and D/H) is likely reflecting the need of proper $\lambda$-scaling. 
To check this, we have added the calculations discussed in the Appendix \ref{sec:Appndx}.

The last two columns use modified $a(T)$ and therefore one has to rely on the numeric
integrations in PRIMAT for $\tau(a)$ and $a(\tau)$. 
That is, we do not have, as far as we know,
analytic SIV solutions for $\tau(a)$ and $a(\tau)$ when $a(T)$ is distorted;
thus, we do instead the PRIMAT numeric integration to get the relevant time variable $\tau(T)$. 
The last column seems to be close to the default PRIMAT run. 
That is because $\lambda$ is close to 1, which is practically PRIMAT
since the last two columns are using the $a_{SIV}(T)$ 
augmented by the distortion factor of $\mathscr{\mathcal{S}}(T)^{-1/3}$
as indicated in column six, or equivalently this is PRIMAT 
$a(T)$ rescaled by $\lambda$ since 
$a_{SIV}/\mathcal{S}^{1/3}=\bar{a}(T)/\lambda$,
where the $\bar{a}(T)$ is the PRIMAT $a(T)$.
The need to have $\lambda$ close to 1 is not an indicator of dark matter content but
indicates the goodness of the standard PRIMAT results that allows only for $\lambda$ close to 1 
as an augmentation, as such leads to a light but important improvement in D/H 
as seen when comparing columns three and seven.

\begin{table}[h]
\small
\begin{tabular}{|c|c|cc|ccccc|}
\hline\hline 
Element & Obs. & PRMT &  $a_{SIV}$ & $\lambda^0$ & $\lambda^{-1}$ & $\lambda^{+1}$ & $\lambda^{1/2}$ & $\lambda^{-1/2}$\\
\hline 
\text{H} & 0.755 & 0.753 & 0.805 & 0.755 & 0.755 & 0.755 & 0.755 & 0.755\\
$Y_{P}\text{=4}Y_{\text{He}}$ & 0.245 & 0.247 & 0.195 & 0.245 & 0.245 & 0.245 & 0.245 & 0.245\\
$\text{D/H}\times10^{5}$ & 2.53 & 2.43 & 0.745 & 2.53 & 2.53 & 2.53 & 2.53 & 2.53\\
\hline 
$^{3}\text{He/H}\times10^{5}$ & 1.1 & 1.04 & 0.746 & 1.05 & 1.03 & 1.08 & 1.07 & 1.04\\
$^{7}\text{Li/H}\times10^{10}$ & 1.58 & 5.57 & 11.9 & 5.24 & 5.66 & 4.81 & 5.02 & 5.45\\
\hline 
$N_{\text{eff}}$ & 3.01 & 3.01 & 3.01 & 3.01 & 3.01 & 3.01 & 3.01 & 3.01\\
$\eta_{10}$ & 6.14 & 6.14 & 6.14 & 1.99 & 1.93 & 2.04 & 2.01 & 1.96\\
$\lambda$ & 1 & 1 & 1. & 1. & 2.43 & 2.76 & 2.66 & 2.5\\
\text{m\v{T}} & 1 & 1 & 1. & 1. & 0.41 & 2.76 & 1.63 & 0.63\\
$\Omega_{b}\; ~ [\%]$ & 4.9 & 4.9 & 4.9 & 1.6 & 1.5 & 1.6 & 1.6 & 1.6\\
$\Omega_{m}~[\%]$ & 31 & 31 & 31 & 5.9 & 7 & 4.8 & 5.3 & 6.4\\
$\sqrt{\chi_{\epsilon}{}^{2}}$ & N/A & 6.8 & 35 & 6.1 & 6.8 & 5.4 & 5.7 & 6.45\\
\hline\hline 
\end{tabular}
\caption{\small{Abundances of the light elements within the standard and SIV BBNS. 
The observational uncertainties are  1.6\% for $Y_{P}$, 1.2\% for D/H, 18\% for He/H, and 19\% for Li/H.
The first five columns are the same as in Table \ref{Table2} to facilitate the easy comparisons with the
remaining columns. $\lambda^{n}$ indicates the chosen scaling \text{m\v{T}}=$\lambda^{n}$  
for the temperature $T^{\beta}$ in the revers reaction formulas.
There is no rescaling of the forward reaction factors  (FRF), nor for the $\exp(\gamma/T_9)$ factor,
or any other temperature dependencies. The last five columns were fitted to reproduce $^{4}$He and D/H.}}
\label{Table3} 
\end{table}
 
In order to study further the SIV guided modifications to the reverse reactions, 
we study only the effects due to \text{m\v{T}} scaling
by utilizing the SIV analytic functions for $a(\tau)$ and $\tau(T)$ 
as in the corresponding middle $a_{SIV}$ columns in Table \ref{Table2}.
The relevant element abundances are given in Table~\ref{Table3}. 
The value of $\lambda$ is set to $1/t_{{\rm in}}$. 
The last five columns were fitted on $\Omega _b$ and $\Omega _m$  
to reproduce $^{4}He$ and $D/H$ since these are known
within 1\% uncertainty. Given that we have chosen  \text{Q/\v{T}}=1; then, 
due to the traditionally preferred energy scaling by $\lambda^{n_m}$ with $n_m=1$, 
the SIV scalings of the thermal energy $k_{B}\,T$ 
with $\lambda$ ($k_{B}=1$) should be $\lambda^{+1}$.
However, the BBNS is in the radiation dominated epoch where, 
$n_T=-1/2$ is expected, as discussed in the Appendix \ref{sec:Appndx}; 
therefore, we have also considered a few other scaling options 
$\lambda^{n}$ with $2n\in\{\pm2,\pm1,0\}$. 
We have also explored the case $n=\pm2$ but
it was not possible to find perfect reproduction for $^{4}He$ and $D/H$ for the
m\v{T}$=\lambda^2$ scaling. That is, we had fit* problem. The minimum was at $\Omega_b=0.02075$ and 
$\Omega_m=0.0863$ with $\lambda=2.26$ and $\sqrt{\chi_{\epsilon}^{2}}=4.8$.
On other hand, m\v{T}$=\lambda^{-2}$ resulted in reproduction of $^{4}He$ and $D/H$ at 
$\Omega_b=0.0149$ and $\Omega_m=0.0809$ with $\lambda=2.31$ and $\sqrt{\chi_{\epsilon}^{2}}=7.5$.
These cases are consistant with what is seen in Table \ref{Table3} but do not have any specific 
justification for choosing such m\v{T}$=\lambda^{\pm2}$ scaling of the $T^\beta$-term in the revers reaction formulas.

From Table \ref{Table3} one can conclude that SIV-guided modifications to the local statistical equilibrium
implemented  to the  $T^\beta$-term in the reverse reactions, which is induced by the \text{m\v{T}}-scaling,
that are consistent with $^4$He and D/H data, are actually \text{m\v{T}}-scaling independent.
The overall result is a reduced baryon and dark matter content in general but no significant 
$\lambda$-scaling dependence.

\section{Summary and Conclusions}\label{sec:Summary}

The SIV analytic expressions for $a(T)$ and $\tau(T)$ can be utilized to study the BBNS within the SIV paradigm.
The functional behavior is very similar to the standard models such as PRIMAT except during the very early universe 
where electron-positron annihilation and neutrino processes affect the $a(T)$ function 
see Table \ref{Table1} and Fig. \ref{tau(a)}. 
The distortion due to these effects encoded in the function $\mathscr{\mathcal{S}}(T)$ 
could be incorporated by considering the SIV paradigm 
as a background state of the universe where the processes could take place.
It has been demonstrated that incorporation of the  $\mathscr{\mathcal{S}}(T)$ 
within the SIV paradigm results in a compatible outcome with the standard BBNS
see Table  \ref{Table2} and if one is to fit the observational data the result is $\lambda\approx1$
for the SIV parameter $\lambda$ (see last column of Table \ref{Table2}). 
However, a pure SIV treatment results in $\Omega_b\approx1\%$  and less total matter,
either around $\Omega_m\approx23\%$ 
when all the $\lambda$-scaling connections are  utilized (see Table  \ref{Table4}), or around 
$\Omega_m\approx6\%$ without any $\lambda$-scaling factors 
(see the fit column of Table  \ref{Table2}).

The SIV paradigm suggests specific modifications to the reaction rates, 
as well as the functional temperature dependences of these rates,
that need to be implemented to have consistence between the 
Einstein GR frame and the WIG (SIV) frame.
In particular, the non-in-scalar factor $T^\beta$ in the reverse reactions rates 
may be affected the most due to the SIV effects. 
As shown in Table \ref{Table3}, we have studied a specific case of dependences and have seen that    
within the assumptions made the SIV model requires three times less baryon matter,
usually around $\Omega_b\approx1.6\%$  and less total matter - around $\Omega_m\approx6\%$.
The lower baryon matter content leads to also a lower photon to baryon ratio $\eta_{10}\approx2$
within the SIV, which is three tines less that the standard value of $\eta_{10}=6.14$.

The results in Table \ref{Table3} indicate insensitivity to the specific 
$\lambda$-scaling dependence of the \text{m\v{T}}-factor in the reverse reaction expressions. 
Thus, one may have to explore further the SIV-guided $\lambda$-scaling relations as done
for the last column in  Table  \ref{Table4}, however, this would require the 
utilization of the numerical methods used by PRIMAT and as such will take us away from the 
SIV-analytic expressions explored in this paper that provide simple model for understanding the BBNS within the SIV paradigm. 
Furthermore, it will take us further away from the accepted local statistical equilibrium and
may require the application of the reparametrization paradigm that seems to  result in SIV like 
equations but does not impose a specific form for $\lambda$ \citep{sym13030379}.

Our main conclusion is that the SIV paradigm provides a concurrent model of the BBNS
that is compatible to the description of $^4$He, D/H, T/H, and $^7$Li/H achieved  in the standard BBNS.
It suffers of the same $^7$Li  problem as in the standard BBNS but also suggests a possible 
SIV-guided departure from local statistical equilibrium which could be a fruitful direction to be explored 
towards the resolution of the $^7$Li  problem.

\section{Appendix: Exploring the SIV-guided $\lambda$-scaling relations}\label{sec:Appndx} 

As mentioned earlier the two scale-factors powers $n_m$ and $n_T$ may not be the same
since one can argue for different scaling powers of the
radiation and rest-mass energies based on the different conservation laws.
For example, in a matter dominated state with $w=0$ one has 
$\rho_m\,a^3\lambda=\rho_{m0}$ with  
$m\propto\rho_m\,a^3\,R_0^3\Rightarrow\,m\rightarrow\,m'_0=m\lambda$,
while for radiation dominated epoch $w=1/3$ one has 
$\rho_r\,a^4\lambda^{2}=\rho_{r0}$, then by using  
$\rho_r\propto\,T^4\Rightarrow\,Ta\lambda^{1/2}=T_0a_0$ along with 
$a\rightarrow\,a'=a\lambda$ this gives $T\rightarrow\,T'=T\lambda^{-1/2}$,
so that the usual $T'a'=T_0a_0$ holds. Thus, while mass scales as $\lambda$
when matter is dominating, then the thermal energy scales\footnote{
This scaling for radiation is consistent with the mass scaling by $\lambda$ 
since $\rho_\gamma\propto\,T^4$ then the total energy in a comoving 3D volume
will be $E_\gamma=\rho_\gamma\,a^3R^3_0\propto\,T^4\,a^3
=T'^4\lambda^{4/2}\,a'^3\lambda^{-3}\propto\,E'_\gamma/\lambda$; 
that is, $E_\gamma\lambda=E'_\gamma$ just as $m\lambda=m'$.
This argument shows that there is no contradiction with the 
law of energy conservation; that is, while the radiation (thermal energy) 
has a different $\lambda$-scaling from the rest-mass energy,
when radiation is absorbed from a finite 3D-region of space the process 
results in the correct energy scaling as for a system with a rest-mass energy,
which is also finite and localized. The key difference is the different $\lambda$-scaling 
of a thermal radiation with a state label T in an infinite volume compared to 
the $\lambda$-scaling of a 3D-localized rest-mass system with a state label m
but consistent upon absorption and emission of localized photons.}
as $\lambda^{-1/2}$ when radiation is dominating. 
Thus, this is the correct thermal energy scaling during the BBNS!
Such $\lambda$-scalings to FRF, m\v{T}, and Q/\v{T}  are easy to be implemented in our SIV study 
during the BBNS where $\lambda$ is practically constant.
In doing so, one has to pull-back the known functions; that is, for a function within the SIV frame
$f(T)$ one has to define its value via the corresponding function $f'(T')$ measured within the EGR laboratory frame.
This way one has $f(T)=f'(T'(T))=f'(T\lambda^{n_T})$ when the two temperatures are related via $T'=T\lambda^{n_T}$.
We will use this to define the $\lambda$-scalings for functions that depend on m\v{T} and Q/\v{T}.
In particular, since our control variable is $T$ then we will adjust only the corresponding scale that comes along
with $T$ but will not include any mass related scaling since the formulas for evaluating $f'(T',m')$ are already using 
EGR laboratory frame values for these quantities; that is, for functions containing m\,T 
the scaling m\v{T}  must be by $\lambda^{-1/2}$ and those that depend on Q/T should be scaled by $\lambda^{+1/2}$.

\begin{table}
\small
\begin{tabular}{|c|cc|ccc|ccc|}
\hline\hline 
Element & Obs. & PRMT & $a_{SIV}$ & fit & fit* & $\bar{a}/\lambda$&  fit* & fit \\
\hline 
 \text{H} & 0.755 & 0.753 & 0.805 & 0.755 & 0.849 & 0.75 & 0.753 & 0.755 \\
 $Y_P=4Y_{\text{He}}$ & 0.245 & 0.247 & 0.195 & 0.245 & 0.151 & 0.25 & 0.247 & 0.245 \\
 $\text{D/H}\times10^5$ & 2.53 & 2.43 & 0.743 & 2.52 & 2.52 & 1.49 & 2.52 & 2.53 \\
\hline  
 $^3\text{He/H}\times10^5$ & 1.1 & 1.04 & 0.745 & 1.05 & 0.825 & 0.884 & 1.05 & 1.04 \\
 $^7\text{Li/H}\times10^{10}$ & 1.58 & 5.56 & 11.9 & 5.24 & 6.97 & 9.65 & 5.31 & 5.42 \\
\hline  
$N_{\text{eff}}$ & 3.01 & 3.01 & 3.01 & 3.01 & 3.01 & 3.01 & 3.01 & 3.01 \\
$\eta_{10}$ & 6.14 & 6.14 & 6.14 & 1.99 & 0.77 & 1.99 & 5.57 & 5.56 \\
 FRF  & 1 & 1 & 1 & 1 & 1.63 & 1 & 1 & 1.02 \\
 \text{m\v{T}} & 1 & 1 & 1 & 1 & 0.78 & 1 & 1 & 0.99 \\
 \text{Q/\v{T}} & 1 & 1 & 1 & 1 & 1.28 & 1 & 1 & 1.01 \\
$\Omega _b$\; [\%] & 4.9 & 4.9 & 4.9 & 1.6 & 0.6 & 1.6 & 4.4 & 4.4 \\
 $\Omega _m$\;[\%]  & 31 & 31 & 31 & 5.9 & 23 & 5.9 & 86 & 95 \\
 $\sqrt{\chi _{\epsilon}^2}$ & N/A & 6.8 & 35 & 6.1 & 14.8 & 22 & 6.2 & 6.4 \\
\hline\hline 
\end{tabular}
\caption{\small{Abundances of the light elements as in Table \ref{Table2} but with two extra columns and 
explicit values of the various reaction parameters (FRF,  \text{m\v{T}},  \text{Q/\v{T}}).
FRF is the forwards rescale factor for all reactions, 
while \text{m\v{T}}  and \text{Q/\v{T}} are the corresponding rescale factors
in the revers reaction formula based on the local thermodynamical equilibrium. 
The SIV $\lambda$-dependences are used when these factors are different from 1;
that is, in the sixth and ninth columns where FRF=$\lambda$, 
\text{m\v{T}}= $\lambda^{-1/2}$, and \text{Q/\v{T}}= $\lambda^{+1/2}$ 
as discussed in the Appendix \ref{sec:Appndx} section. 
The columns denoted by fit contain the results for perfect fit 
on $\Omega _b$ and $\Omega _m$  to $^{4}$He and D/H,
while fit* is the best possible fit on $\Omega _b$ and $\Omega _m$ 
to the $^{4}$He and D/H observations 
for the model considered as indicated in the columns four and seven.
The middle three columns (four to six) depict results produced by 
using the analytical functions within a ``naive'' SIV model without modification of the reaction rates,
while the case of SIV suggested reaction rates modifications is shown in the sixth column.
The last three columns are usual PRIMAT runs with modified $a(T)$ 
such that $\bar{a}/\lambda=a_{SIV}/\mathcal{S}^{1/3}$,
where $\bar{a}$ is the PRIMAT's $a(T)$ for the decoupled neutrinos case.
Column seven is actually $a_{SIV}/\mathcal{S}^{1/3}$,
but it is denoted by $\bar{a}/\lambda$ to remind us 
about the relationship $a'=a\lambda$;
the run is based on $\Omega _b$ and $\Omega _m$ from column five.
The different values of $\eta_{10}$ are due to different values of
$h^2\Omega_b$, which can be seen by noticing that 
$\eta_{10}/\Omega_b$ is always $\approx 1.25$.}}
\label{Table4} 
\end{table}

The SIV runs shown in Table \ref{Table2} could be viewed as a ``naive'' SIV model because 
we have not utilized any $\lambda$-modifications of the 
various reaction parameters (FRF,  \text{m\v{T}},  \text{Q/\v{T}}).
The study of using such modifications are presented in column six and nine of Table \ref{Table4},
while the other columns are the same as  in Table \ref{Table2}. 
The fit* here is the best possible, but not perfect, fit for  $^4$He and D/H and 
seems to require a reduced total mass content relative to PRIMAT 
but much more than in the previous case (column five).
The failure of this fit* (column six) to achieve perfect fit to the 
$^4$He and D/H reflects the importance of the electron-positron annihilation process
accounted by $\mathscr{\mathcal{S}}(T)$.

We have already discussed the contribution of $\mathscr{\mathcal{S}}(T)$ within SIV
as shown in  Table \ref{Table2} which uses a simple (``naive'') SIV model without the utilization of the 
$\lambda$-modifications of the various reaction parameters (FRF,  \text{m\v{T}},  \text{Q/\v{T}}). 
We already suggested that, the second fit* failure to achieve a perfect 2D fit (on $^4$He and D/H) 
is likely reflecting the need for such $\lambda$-modification implementation. 
To check this, we have performed the calculations shown in the last column in Table \ref{Table4}.

The closeness of  $\Omega _m$ to 1 in the last two columns is actually reflecting the 
need of having FRF close to 1.
This is because FRF is present in the thermonuclear reactions
as well as in the weak reactions where the time scale is set naturally by the neutron life-time.
Thus, in order to not change the weak reactions significantly, 
which are related to the neutron life-time, one has to have FRF close to 1.
Since, FRF is expected to be equal to $\lambda$,
this leads to $\lambda$ close to 1 as well. Thus, we kept FRF=1 in the runs for the two 
columns shown that use modified $a(T)$ and therefore one has to rely on the numeric
integrations in PRIMAT for $\tau(a)$ and $a(\tau)$. That is because we do not have
analytic SIV solutions for $\tau(a)$ and $a(\tau)$ when $a(T)$ is distorted;
thus, we do the numeric integration instead. 
The last column seems to be close to the default PRIMAT run. 
That is because $\lambda$ is close to 1, which is practically PRIMAT
since the last three columns are using the $a_{SIV}(T)$ 
augmented by the distortion factor of $\mathscr{\mathcal{S}}(T)^{-1/3}$
as indicated in column six, or equivalently this is PRIMAT 
$a(T)$ rescaled by $\lambda$ since 
$a_{SIV}/\mathcal{S}^{1/3}=\bar{a}(T)/\lambda$,
where the $\bar{a}(T)$ is the PRIMAT $a(T)$.
The need to have $\lambda$ close to 1 is not an indicator of dark matter content. 
These results indicate that unmodified FRF (=1) is preferred,
which pushes $\lambda$ towards 1 when considered as a possible modification option
as the fit in the last column indicates. For this fit we allowed the 
$\lambda$-scaling for m\v{T} and Q/\v{T} to depend on $\lambda$ as
stated in the table caption and discussed above but we got back
to the PRIMAT case with a little smaller  $\Omega _b$ and  
$\Omega _m$ very close to 1 since this is an effective way of getting $\lambda=1$.


\begin{thebibliography}{99}


\bibitem[Bondi(1990)]{Bondi90} Bondi, H. 1990, in Modern Cosmology
in Retrospect, Eds. {Bertotti}, B., {Balbinot}, R., \& {Bergia}, S.,{Cambridge Univ. Press.}, 426 pp.

\bibitem[{{Bouvier} \& {Maeder}(1978)}]{BouvierM78}
{Bouvier}, P. \& {Maeder}, A. 1978, \apss, 54, 497

\bibitem[Bronstein \& Semendiaev(1974)]{Bronstein74} 
Bronstein, L.N., Semendiaev, K.A. 1974, Aide-memoire de mathematiques, 
Ed. Eyrolles, Paris, 935 p.

 
\bibitem[{{Canuto} {et~al.}(1977){Canuto}, {Adams}, {Hsieh}, \& {Tsiang}}]{Canu77}
{Canuto}, V., {Adams}, P.~J., {Hsieh}, S.-H., \& {Tsiang}, E. 1977, \prd, 16, 1643



\bibitem[{{Carroll} {et~al.}(1992){Carroll}, {Press}, \& {Turner}}]{Carr92}
{Carroll}, S.~M., {Press}, W.~H., \& {Turner}, E.~L. 1992, \araa, 30, 499

 

\bibitem[Carter(1979)]{Carter79} 
Carter, B. 1979, in ''Confrontation of cosmological theories with observational data'', 
IAU Symp. 63, Reidel Publ. Co., Dordrecht, p. 291.

\bibitem[Coles \& Lucchin(2002)]{Coles02} 
Coles, P., Lucchin, F. 2002, Cosmology. The Origin and Evolution of Cosmic Structure,
Wiley \& Sons Ltd, 492 p.


\bibitem[{{Dirac}(1973)}]{Dirac73}
{Dirac}, P.~A.~M. 1973, Proceedings of the Royal Society of London Series A,
 333, 403

\bibitem[Durrer(2008)]{Durrer08} 
Durrer, R. 2008, The Cosmic Microwave Background, Cambridge Univ. Press, 401 p.


 


\bibitem[Jesus(2018)]{Jesus18} Jesus, J.F. 2018, arXiv:1712.00697


\bibitem[Maeder(2017a)]{Maeder17a} Maeder, A. 2017a, \apj, 834, 194 

 \bibitem[Maeder(2017b)]{Maeder17b} Maeder, A. 2017b, \apj, 847, 65
 
 \bibitem[Maeder(2017c)]{Maeder17c} Maeder, A. 2017c, \apj, 849, 158
 
 \bibitem[Maeder(2018)]{Maeder18} Maeder, A. 2018, A. arXiv:1804.04484
 
 \bibitem[Maeder \& Bouvier (1979)]{MBouvier79} Maeder, A., Bouvier, P. 1979, Astron. Astrophys., 73, 82
 

\bibitem[Maeder \&  Gueorguiev(2019)]{MaedGueor19} 
Maeder, A.; Gueorguiev, V.G. The growth of the density fluctuations in the scale-invariant vacuum theory. 
{\em Phys. Dark Univ.} {\bf 2019}, {\em 25}, 100315. 

\bibitem[Maeder  \& Gueorguiev(2020)]{MaedGueor20a} 
Maeder, A.; Gueorguiev, V.G. The Scale-Invariant Vacuum (SIV) Theory:  A Possible Origin of Dark Matter and Dark Energy. 
{\em Universe} {\bf 2020}, {\em 6}, 46. 

\bibitem[Maeder \&  Gueorguiev(2020)]{MaedGueor20b} 
Maeder, A.; Gueorguiev, V.G. Scale-invariant dynamics of galaxies, MOND, dark matter, and the dwarf spheroidals. 
{\em MNRAS} {\bf 2019}, {\em 492}, 2698.

\bibitem[Maeder \&  Gueorguiev(2021)]{SIV-Inflation'21}
Maeder, A.; Gueorguiev, V.G. Scale invariance, horizons, and inflation.
{\em MNRAS} {\bf 2021}, {\em 504}, 4005.

\bibitem[Gueorguiev \&  Maeder(2021)]{univ8040213}
Gueorguiev, V.G.; Maeder, The Scale Invariant Vacuum Paradigm: Main Results and Current Progress.
{\em Universe} {\bf 2022}, {\em 8}, 213. 

\bibitem[Gueorguiev \&  Maeder(2021)]{sym13030379}
Gueorguiev, V.G.; Maeder, A. Geometric Justification of the Fundamental Interaction Fields for the Classical Long-Range Forces.
{\em Symmetry} {\bf 2021}, {\em 13}, 379.

\bibitem[Mukhanov(2004)]{Mukhanov04} Mukhanov, V. 2004, 
Intnl. J. Theoretical Physics, 43, 669

\bibitem[Steigman(2007)]{Steigman07} Steigman, G. 2007, Ann. Rev.
Nuclear and Particle Sci. 57, 463

\bibitem[Weinberg(2008)]{Weinberg08} Weinberg, S. 2008, Cosmology,
Oxford Univ. press, 593 p.

 
\bibitem[Weyl(1923)]{Weyl23}
{Weyl}, H. 1923, Raum, Zeit, Materie. Vorlesungen {\"u}ber allgemeine
 Relativit{\"a}tstheorie. Re-edited by Springer Verlag, Berlin, 1970


\bibitem{Maeder18} 
A. Maeder, ``Evolution of the early Universe in the scale invariant theory," ArXiV: 1902.10115

\bibitem{PRIMAT} 
Pitrou, C., Coc, A., Uzan, J.-P., Vangioni, E. 
``Precision big bang nucleosynthesis with improved Helium-4 predictions.''\\
Physics Reports 754, 1–66 (2018). 
ArXiV: 1801.08023 

\end{thebibliography}
\end{document}